# Highly-stable, flexible delivery of microjoule-level ultrafast pulses in vacuumized anti-resonant hollow-core fibers for active synchronization

CHUANCHUAN YAN,[1,2,†] HONGYANG LI,[2,3,†] ZHIYUAN HUANG,[2,7] XINLIANG WANG,[2] DONGHAN LIU,[2] XINGYAN LIU,[2] JINYU PAN,[2] ZHUOZHAO LUO,[1,2,4] FEI YANG,[5] YU ZHENG,[6] RUOCHEN YIN,[6] HAIHU YU,[1] YUXIN LENG,[2] LIWEI SONG,[2,8] MENG PANG,[2,4] AND XIN JIANG,[1,4,*]

[1]*National Engineering Research Center of Fiber Optic Sensing Technology and Networks, Wuhan University of Technology, Wuhan 430070, China*
[2]*State Key Laboratory of High Field Laser Physics and CAS Center for Excellence in Ultra-intense Laser Science, Shanghai Institute of Optics and Fine Mechanics (SIOM), Chinese Academy of Sciences (CAS), Shanghai 201800, China*
[3]*School of Physics Science and Engineering, Tongji University, Shanghai 200092, China*
[4]*Russell Centre for Advanced Lightwave Science, Shanghai Institute of Optics and Fine Mechanics and Hangzhou Institute of Optics and Fine Mechanics, Hangzhou 311400, China*
[5]*Key Laboratory of Space Laser Communication and Detection Technology, Shanghai Institute of Optics and Fine Mechanics, Chinese Academy of Sciences, Shanghai 201800, China*
[6]*IFiber (Ningbo) Optoelectronics Technology Co., LTD., Ningbo 315000, China*
[7]*e-mail: huangzhiyuan@siom.ac.cn*
[8]*e-mail: slw@siom.ac.cn*
*Corresponding author: jiangx@whut.edu.cn; Current e-mail: xin.jiang@r-cals.com*
[†]*These authors contributed equally to this letter.*

**We demonstrate the stable and flexible light delivery of multi-μJ, sub-200-fs pulses over a ~10-m-long vacuumized anti-resonant hollow-core fiber (AR-HCF), which was successfully used for high-performance pulse synchronization. Compared with the pulse train launched into the AR-HCF, the transmitted pulse train out of the fiber exhibits excellent stabilities in pulse power and spectrum, with pointing stability largely improved. The walk-off between the fiber-delivery and the other free-space-propagation pulse trains, in an open loop, was measured to be <6 fs root-mean-square (RMS) over 90 minutes, corresponding to a relative optical-path variation of <2×10$^{-7}$. This walk-off can be further suppressed to ~2 fs RMS simply using an active control loop, highlighting the great application potentials of this AR-HCF set-up in large-scale laser and accelerator facilities.**

Timing distribution technique, based on active synchronization of ultrafast optical pulses, has been widely used in a variety of large-scale laser and accelerator facilities, providing unprecedented temporal precision [1-5]. In the PW-level, ultra-intense laser facility [6], for example, high-quality timing synchronization of optical pulses plays a critical role in the pump-probe experiment, guaranteeing sufficient pulse overlaps and interactions [3]. One core requirement of a timing distribution system is to deliver stably ultrafast pulses over long distances. While optical fibers are widely recognized as an ideal medium for stable and flexible light-wave transmission [5,7,8], high-performance delivery of μJ-level ultrafast pulses generated from Ti: sapphire laser system over conventional silica-core fiber has proved difficult to achieve. The ultra-high peak power of such high-energy, ultrafast pulses quickly causes excessive nonlinearity accumulation in the silica fiber core, leading to largely-distorted pulse structures and optical spectra. Due to this limitation, previous fiber-based timing distribution systems typically operate at pulse energies of sub-nJ levels, with pulse widths ranging from several hundreds of fs to several ps [5,9]. Even though some attempts have recently been performed to use anti-resonant hollow-core fibers (AR-HCFs) for nonlinear-free delivery of high-energy pulses [10], systematic studies on the performances of such hollow-core fibers (HCFs) for ultrafast pulse synchronization have not been reported so far.

In this letter, we demonstrate an active synchronization set-up of ultrafast laser pulses in which a 9.6-m length of vacuumized AR-HCF provides the high-performance flexible delivery of μJ-level, sub-200 fs pulses. In the vacuum core of the fiber, nonlinear spectral broadening effect was utterly suppressed, while the pulse temporal width decreased slightly due to small dispersion of the hollow waveguide. Without dispersion compensating or auto-alignment element, the output pulse train from the AR-HCF exhibited perfect stabilities in both optical spectrum and average power, with pointing stability improved from ~6 μrad to <1 μrad. Using a balanced optical cross-correlator (BOC) set-up, we measured the long-term walk-off between this HCF-delivered optical beam and the other one of free-space propagation in the laboratory environment, the results gave a root-mean-square (RMS) timing jitter of 5.23 fs over 90 minutes, corresponding to a small relative path variation of 0.2 parts per million. This open-loop path variation was further suppressed to less than 0.1 parts per million using an active control loop, leading to a timing-synchronization precision of ~2 fs RMS.

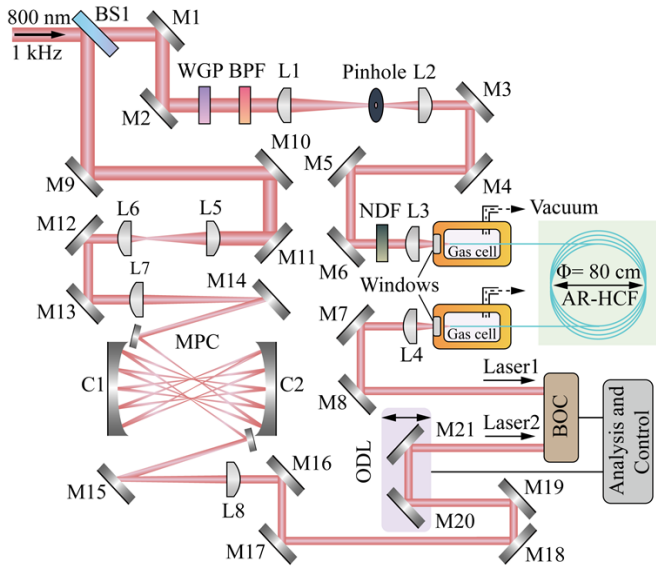

Fig. 1. Experimental set-up. BS1, beam splitter; M1-M21, silver mirrors; WGP, wire grid polarizer; BPF, bandpass filter; L1-L8, plano-convex lens; C1-C2, concave mirrors; NDF, neutral density filter; AR-HCF, anti-resonant hollow-core fiber; MPC, multi-pass cavity; BOC, balanced optical cross-correlator; ODL, optical delay line.

The set-up used in the experiment is sketched in Fig. 1. The pulses were from a high-energy optical parametric amplifier Ti: sapphire laser system, with 800 nm central wavelength, ~35 fs pulse width, ~470 mW average power and 1 kHz repetition rate. As shown in Fig. 1, the pulse train passed through a 10:90 beam splitter (BS1), and energies of the transmitted (90%) pulses were controlled by a wire grid polarizer (WGP). A bandpass filter (BPF) with a 10-nm bandwidth at 800 nm was employed to narrow down the pulse spectrum, leading to a broadening of the temporal pulse width from ~35 fs to ~200 fs. The spectral narrowing of pulses can efficiently suppress the temporal broadening of the pulses over AR-HCF propagation due to fiber dispersion. In order to improve the beam quality of the pulse train, see Figs. 2(b) and 2(c), a spatial filter consisting of a 150 μm pinhole and two coated plano-convex lenses with focal lengths of 1 m (L1) and 0.4 m (L2) was used in the set-up.

The pulse train through the spatial filter was then launched into a 9.6-m length of vacuumized AR-HCF using a coated plano-convex lens (7.5 cm focal length, L3). The energies of the input pulses can be adjusted using a variable neutral density filter (NDF). In the experiment, the AR-HCF was coiled with a diameter of ~80 cm, and the two ends of the AR-HCF were mounted inside two gas cells, both of which were connected to a vacuum pump (ISP-250C, IWATA) and sealed with 5-mm-thick fused silica windows. The scanning electron micrograph (SEM) of the AR-HCF is illustrated in the inset figure of Fig. 2(a), with a core diameter of ~30 μm and a wall thickness of ~520 nm. In Fig. 2(a), the loss profile (solid red line) of the AR-HCF was measured using the cut-back method, giving loss values of ~0.27 dB/m at 800 nm (blue dashed line) and ~0.11 dB/m at 1550 nm (purple dashed line). The light green bar indicates the fiber loss in the wavelength range from 790 to 860 nm. The coupling efficiency into the AR-HCF was measured to be ~85%. When the input pulse energy was 8 μJ (8 mW average power), the pulse energy output from the AR-HCF was measured to be 3.4 μJ. The total insertion loss of the fiber-delivery section was ~3.7 dB, and the near-field beam profile of the output pulse train is shown in Fig. 2(d).

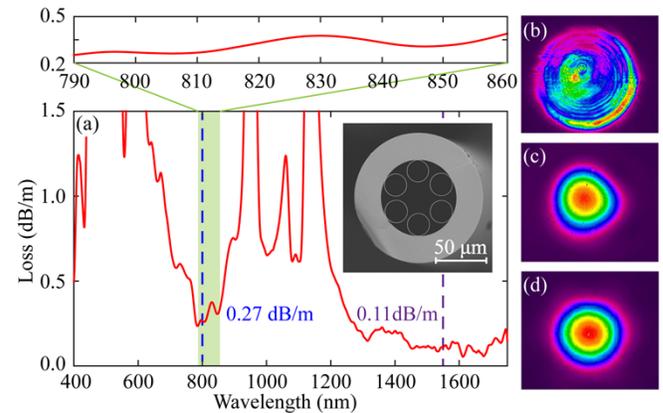

Fig. 2. (a) Measured loss (solid red line) of AR-HCF, blue and purple dashed lines indicate the fiber loss at 800 nm and 1550 nm, which are 0.27 dB/m and 0.11 dB/m, respectively. The light green bar indicates the fiber loss in the wavelength range from 790 to 860 nm. The inset represents the SEM of AR-HCF with a core diameter of 30 μm. (b, c, d) Near-field beam profiles, (b) before BS1, (c) and (d) at the input and output of the AR-HCF.

In the experiment, we gradually increased the pulse energy launched into the AR-HCF from 2 to 8 μJ, and measured the input and output pulse spectra at different pulse energies. The results are plotted in Fig. 3. Without the vacuum pumping, the hollow-core of the fiber was filled with ~1-bar air. At μJ-level pulse energies, the gas nonlinearity is strong enough to cause obvious pulse spectral broadening and distortion, as shown in Fig. 3(a). As Raman-active gases, the oxygen and nitrogen molecules in air can lead to pulse wavelength redshifts, see Fig. 3(a). Due to the accumulated Raman effect, the redshifted wavelength increased from ~836 to ~880 nm as the pulse energy increased from 2 to 8 μJ. Note that the intensity of the Raman-shifted component gradually decreases as it moves to longer wavelengths at higher pulse energies, mainly due to the higher fiber loss at these longer wavelengths [see Fig. 2(a)]. At pulse energies higher than 2 μJ, besides the Raman-shifted component, several spectral spikes with relatively-lower intensities were observed on the output pulse spectra, see Fig. 3(a). This is mainly because the pulse underwent soliton self-compression and splitting due to the gas nonlinearity [11]. Such nonlinearity-induced pulse distortions can be efficiently suppressed when the hollow-core of the fiber is vacuumized. In Fig. 3(b), the pulse spectra output from the AR-HCF at pulse energies up to 8 μJ remain almost unchanged.

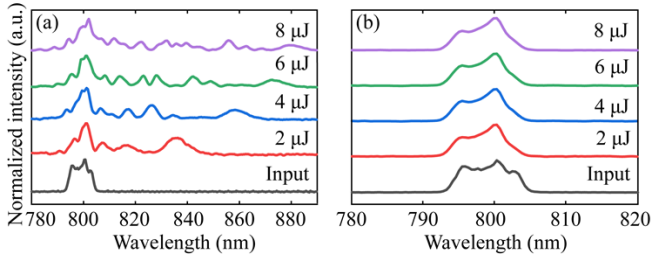

Fig. 3. The spectral evolutions out of the AR-HCF as a function of input pulse energy. (a) Air-filled AR-HCF, (b) vacuumized AR-HCF. The solid black lines in (a) and (b) represent the input pulse spectrum.

These stable optical spectra indicate clean temporal profiles of the output pulses which have been verified in our experiments. Using a second-harmonic-generation frequency-resolved optical gating (SHG-FROG) set-up, we measured the pulse temporal profiles before and after the AR-HCF at an input pulse energy of 8 μJ, and the results are plotted in Fig. 4. The retrieved FROG traces show good agreement with the measured ones [see Figs. 4(a), 4(b), 4(d) and 4(e)], and the retrieved pulses have full-width-half-maximum widths of 197 fs and 167 fs respectively [see Figs. 4(c) and 4(f)]. This slight decrease of pulse width could be interpreted as: the input pulse of 197 fs has some positive chirp which was partially compensated by the anomalous dispersion of the hollow-core fiber.

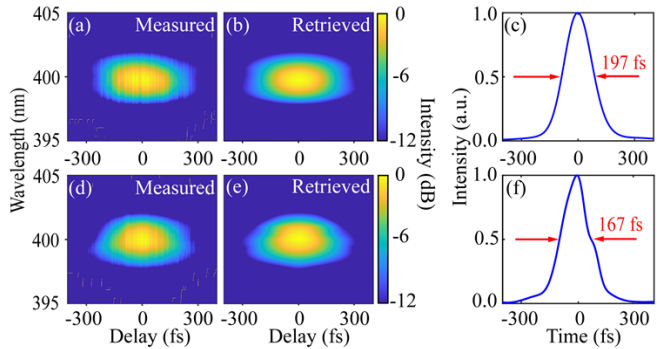

Fig. 4. Time-domain characterization of the pulses at the input (a-c) and output (d-f) of AR-HCF. (a, d) Measured SHG-FROG traces. (b, e) Retrieved SHG-FROG traces. The FROG errors for both cases are ~0.3%. (c, f) Retrieved temporal profiles.

In the experiment, we also measured respectively the power, pointing and spectral stabilities of pulse trains at both the input and output ends of the AR-HCF. The results are illustrated in Fig. 5. As shown in Figs. 5(a) and 5(d), when the input power was 5.16 mW, the output power was 2.16 mW. In order to evaluate the power fluctuations, we recorded the laser average power over 30 minutes and calculated the coefficient of variation, which is defined as the ratio of standard deviation to the average value, to be 0.998% and 1.009% for respectively the input and the output pulse trains. The almost identical power-fluctuation values imply the capability of the AR-HCF for robust delivery of high-energy ultrafast pulses.

Then, we focused the input and output pulse trains on the CCD camera using a lens with 10-cm focus length to measure pointing stabilities of the two beams within 30 minutes. As shown in Fig. 5(b), the input pulse train exhibits pointing stability of $\theta_x$ = 5.87 μrad, $\theta_y$ = 6.92 μrad at two orthogonal directions. These stability values were largely improved at the output port of the AR-HCF, with measured values of $\theta_x$ = 0.89 μrad, $\theta_y$ = 0.92 μrad [see Fig. 5(e)], highlighting one key advantage of this fiber waveguide for highly-stable light delivery.

The measured spectra of the input and output pulse trains are shown in Figs. 5(c) and 5(f). For each measurement, we collected 600 recordings of pulse spectrum, and the integration time of each recording was 1s. We can see from Figs. 5(c) and 5(f) that the spectrum of the output pulse train exhibits a bit noisier than that of the input, which we believe results mainly from the higher relative noise level of the fiber-based spectrometer (HR2000+, Ocean Optics) when measuring output pulse train with a lower average power. In the experiment, in order to avoid possible damage to the spectrometer, only a small portion of the ultrafast pulses was coupled into the spectrometer, leading to relatively-low signal-to-noise ratios of the measurements.

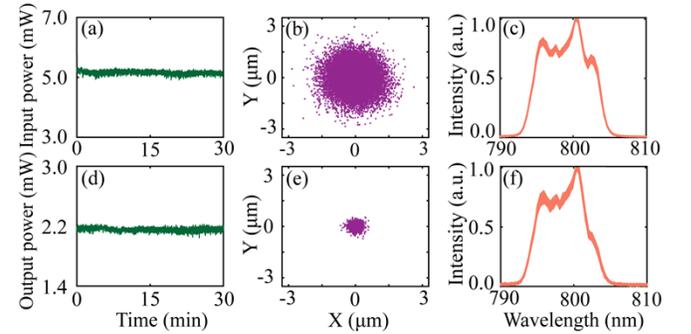

Fig. 5. The power stability, pointing stability, and spectral stability of the pulse at the input (a-c) and output (d-f) of AR-HCF. (a, d) Power stability, (b, e) pointing stability, (c, f) spectral stability.

To verify the high stability of the pulse delivery over the AR-HCF, in the experiment we constructed the second free-space path of ultrafast pulse propagation. As illustrated in Fig. 1, the reflected beam from BS1 with 10% energy passed through two coated lenses with focal lengths of 30 cm (L5) and 10 cm (L6) for beam shaping, and then was launched into a multi-pass cavity (MPC) using a coated lens with a focal length of 1 m. The MPC consists of two concave mirrors with focal lengths of 25 cm, which are coated with high reflectivity films with high reflection coefficients of 98.5%. The pulse train traveled back and forth 5 times in this 0.8-m-long MPC, leading to an effective propagation distance of ~8 m. The pulse train from the MPC passed through an optical delay line (ODL), and then was collimated by a coated lens with a focal length of 50 cm (L8). The total propagation distance of the second path was adjusted to be ~10 m so as to match that of the AR-HCF path. The two laser beams [Laser1 and Laser2 in Fig. 6(a)] from the two paths were finally launched into the BOC module [12] for timing jitter measurements.

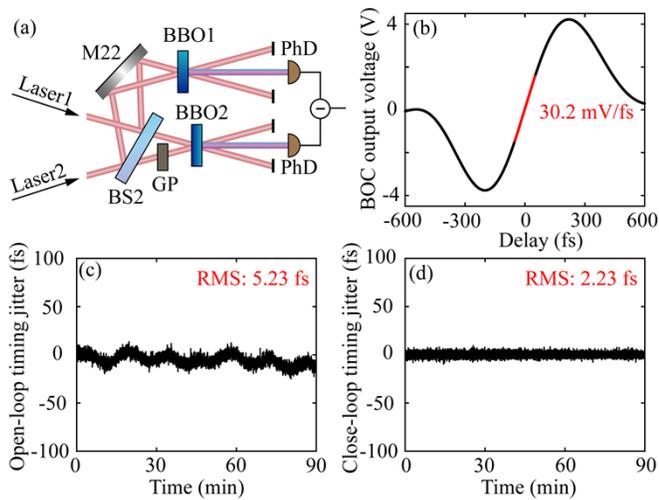

Fig. 6. (a) The experimental set-up of BOC. BS2, beam splitter; M22, silver mirror; GP, glass plate; BBO1-BBO2, beta-barium borate; PhD, photodetector. (b) The optical cross-correlation traces of the BOC (solid black line). The solid red line indicates the fitting slope with 30.2 mV/fs. (c) Open-loop timing jitter. (d) Close-loop timing jitter.

As shown in Fig. 6(a) and described in [12], the BOC module based on sum-frequency generation, was used to measure directly timing jitter between the two pulse trains. When a 200-μm-thick glass plate (GP) and two type-I BBO crystals with the same thickness of 2 mm were used in the BOC set-up, the resulting signal slope was measured to be 30.2 mV/fs, see Fig. 6(b). As the average powers of the two pulse trains launched into the BOC module were adjusted to be 0.6 mW, under the open-loop condition, the RMS timing jitter value was measured to be 5.23 fs over 90-min recording [see Fig. 6(c)], corresponding to very low relative optical-path-variation of $<2\times10^{-7}$ in the laboratory environment. This relative path variation is more than one order of magnitude lower than that in conventional silica-core fiber [13,14] due to the ultra-low light-silica overlap of the guided optical mode in the AR-HCF [15]. Most of light fields propagating in this AR-HCF are tightly confined in vacuum by the anti-resonance guiding mechanism, making the light wave in this AR-HCF rather insensitive to temperature fluctuations and fiber micro-bending [8]. The RMS timing jitter can be easily suppressed to 2.23 fs using the close-loop control based on the ODL set-up, see Figs. 1 and 6(d).

In conclusion, we demonstrated the flexible and stable delivery of μJ-level, sub-200-fs Ti: sapphire laser pulses over 9.6-m-long vacuumized AR-HCF. With vacuum pumping, the effects due to fiber nonlinearity were sufficiently suppressed, leading to almost constant pulse spectra over transmission. The pulse width at the fiber output port decreased by ~15% due to the waveguide dispersion, which has negligible influence on the following active pulse-synchronization experiments. The power, pointing and spectral stability of the pulse train after AR-HCF delivery were carefully measured, giving remarkably-high performances of ~1% power fluctuation, <1 μrad pointing stability and robust spectral stability. BOC measurements exhibit RMS timing jitter of 5.23 fs (open-loop) and 2.23 fs (close-loop) for this AR-HCF pulse delivery system, highlighting its excellent immunity to environmental perturbations.

**Funding.** Zhangjiang Laboratory Construction and Operation Project (20DZ2210300); National Natural Science Foundation of China (52072277); CAS Project for Young Scientists in Basic Research (YSBR-059); Basic research project of Shanghai Science and Technology Innovation Action Plan (20JC1416000); National Postdoctoral Program for Innovative Talents (BX2021328); National Natural Science Foundation of China Youth Science Foundation Project (62205353); China Postdoctoral Science Foundation (2021M703325).